\documentclass[prl,twocolumn,showpacs,aps,floatfix,amsmath,amssymb,amsfonts]{revtex4-1}
\pdfoutput=1
\usepackage{amsmath}
\usepackage{amsfonts}
\usepackage{graphicx}
\usepackage{subfigure}
\usepackage[T1]{fontenc}
\usepackage{color}

\begin{document}
\title{Fast Quantum Gate via Feshbach-Pauli Blocking in a Nanoplasmonic Trap}
\author{Krzysztof Jachymski$^{1,2}$, Zbigniew Idziaszek$^{1}$ and Tommaso Calarco$^{2}$}
\affiliation{$^1$Faculty of Physics, University of Warsaw, Ho{\.z}a 69,
00-681 Warsaw, Poland,\\
$^2$Institut f\"{u}r Quanteninformationsverarbeitung, Universit\"{a}t Ulm, D-89069 Ulm, Germany}
\date{\today}

\begin{abstract}
We propose a simple idea for realizing a quantum gate with two fermions in a double
well trap via external optical pulses without addressing the atoms individually. The
key components of the scheme are Feshbach resonance and Pauli blocking, which
decouple unwanted states from the dynamics. As a physical example we study atoms in
the presence of a magnetic Feshbach resonance in a nanoplasmonic trap and discuss
the constraints on the operation times for realistic parameters, reaching a fidelity
above $99.9\%$ within $42\mu$s, much shorter than existing atomic gate schemes.
\end{abstract}

\maketitle
Quantum gates are the building blocks for quantum simulation and computation~\cite{Nielsen2010}. Quantum computing can be fault-tolerant only if
elementary gate errors are kept below a given threshold~\cite{Preskill1998}, estimated
to be from around $10^{-3}$ to $10^{-2}$ depending on the type of
error correction~\cite{knill,hollenberg}. Error sources are
twofold: firstly, the finite decoherence time of quantum bits implies a
loss of coherence during the gate operation; secondly, dynamical imperfections (imprecise
experimental control or 'leakage' to states outside the computational subspace)
limit the intrinsic fidelity of the gate. The goal of quantum gate
schemes is to minimize the ratio between the operation time and the qubits'
decoherence time, while maximizing the gate fidelity.

A number of different systems, including ions, photons and superconducting
circuits, are suitable for building a quantum gate. Among the promising candidates
are ultracold neutral atoms and molecules~\cite{CiracZoller, Cirac2000,
Benhelm2008,Milburn,Burkard1999,Jaksch1999,Deutsch1999,Jaksch2000,DeMille2002,Calarco2004,negretti2011}.
The main reasons for this are long qubit coherence times and accurate control of the
dynamics, resulting in low error rates, achievable in those
systems~\cite{BlochRMP,Anderlini2007}. Furthermore, they can be arranged in regular
arrays using optical lattices~\cite{Jaksch1998,Greiner2002,Bloch2005}, which provides
unparalleled scalability and are amenable to operations addressing each qubit
individually~\cite{Wurtz,Bakr2009,Weitenberg2011}. The main problem with such
systems is that achievable gate speeds are comparatively low due to weak
atomic interactions and limited trapping frequencies.

A powerful tool to control the interactions of trapped neutral atoms is
provided by Feshbach resonances, which allow for manipulation of the scattering
length~\cite{JulienneRMP,Inouye,Bartenstein2004}. The resonance mechanism comes from the coupling of
a free pair of atoms with a molecular bound state. The energy of the bound state can
be controlled by an external field and the resonance occurs when it crosses the threshold
energy of the open channel.

Recently a new kind of traps for ultracold atoms has been
proposed~\cite{Murphy2009,Chang2009,Zoller2012,Julia2013}, allowing to reach
significantly higher trapping frequencies than before. Experiments with atoms
coupled to nanostructures are developing quickly~\cite{Stehle2011,Lukin2013}. In
these novel systems, the atoms can be trapped by the laser light scattered on plasmonic nanostructures. The resulting
trapping frequencies can be of the order of several MHz. 

\begin{figure}[b]
\includegraphics[width=0.33\textwidth]{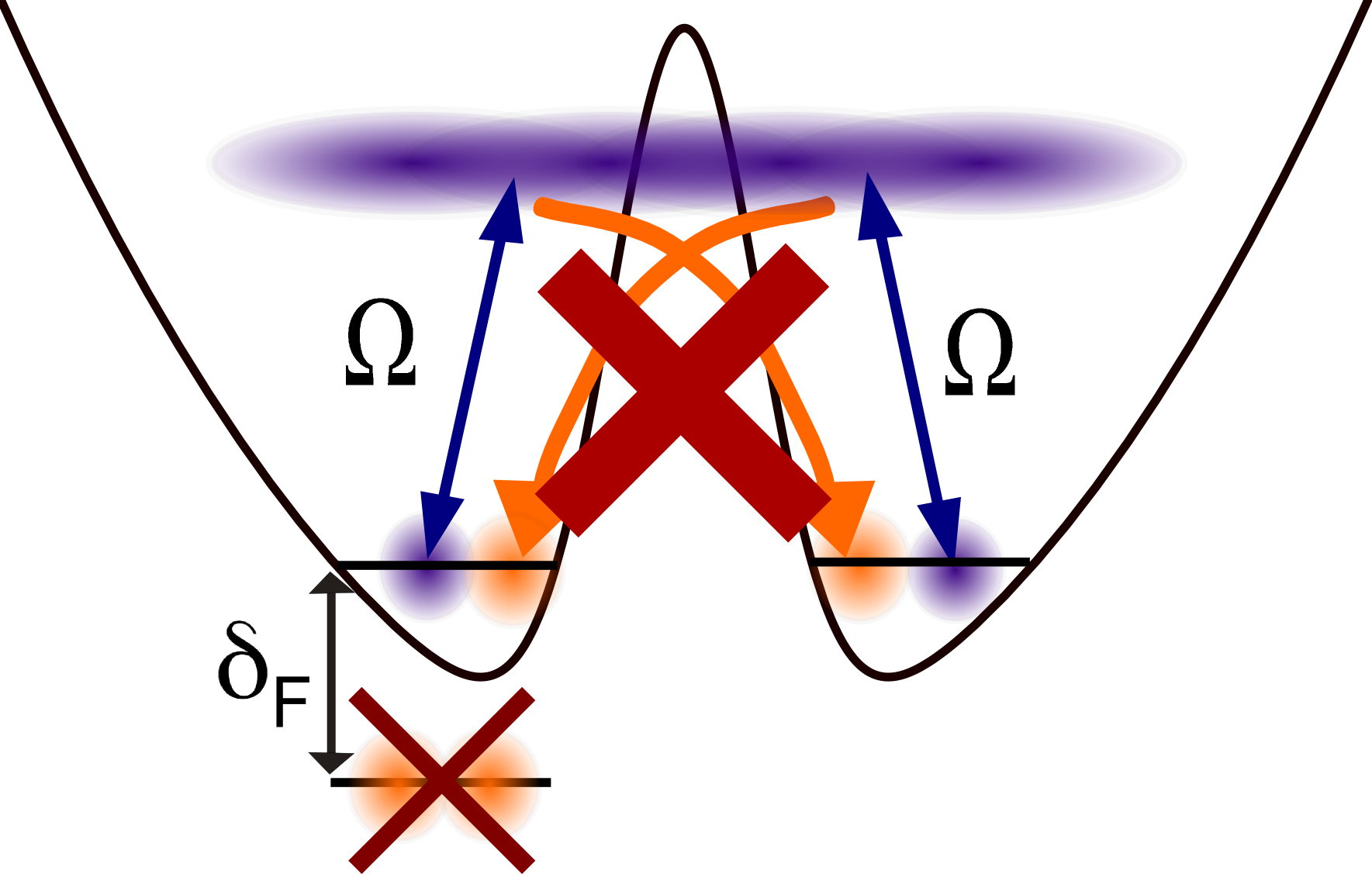}
\caption{\label{setup}(color online) Schematic view of the gate operation. Two
particles (purple balls) are initially localized in separated wells of the trap,
then an external pulse with Rabi frequency $\Omega$ (blue arrows) is applied to
induce coupling to an excited, delocalized trap state. The state with doubly
occupied site is detuned by $\delta_F$ due to the Feshbach resonance when the atoms
are in antisymmetric internal state or forbidden by Pauli blocking when they are in
symmetric internal state (crossed orange arrows and balls).}
\end{figure}
The combination of strong resonant interactions with tight subwavelength traps
gives an excellent opportunity to implement much faster quantum gates.
However, there is no way to achieve single site addressability in those
systems. Existing proposals circumventing this problem~\cite{Hayes2007} do not provide significant improvement, as the traps
need to be moved adiabatically with respect to the small tunnelling energy splitting.

In this Letter we propose to add a missing ingredient allowing for significant speedup. We show that by combining resonant atomic interactions with Pauli blocking we can achieve a fast gate while keeping gate errors below $10^{-3}$ in a realistic setting without individual qubit addressing. Our idea is schematically presented on Figure~1. We consider a pair of spin-$1/2$ fermions in a tight double well trap and an external field inducing transitions between different trap levels. The time evolution of the pair depends on the spin state due to symmetry requirements. The Feshbach resonance plays a crucial role here as a mechanism for suppressing unwanted states in the evolution. Together with Pauli exclusion principle, it blocks transitions to states with both atoms in the same trap level. The lack of separation of center of mass and relative motion has to be included in the calculations~\cite{Kestner2010,Buchler,Sala2012,Jachymski}. The gate fidelity can be enhanced by means of optimal control~\cite{CRAB}. We estimate experimentally realistic gate times in our scheme to be of the order of tens of microseconds, more than an order of magnitude faster than the best current schemes using neutral atoms~\cite{Sherson2014}.

\textit{Double well trap.}
Let us consider a pair of neutral atoms trapped by the light scattered on a set of nanostructures~\cite{Murphy2009,Chang2009}. We assume that the trap is designed to produce a double well potential. For simplicity we describe it as a one-dimensional harmonic potential with frequency $\omega$ with a Gaussian barrier in the middle, parametrized by width $d$ and height $b$. Submicrometer precision of arranging the nanostructures and tunability of polarization of light sources~\cite{Stehle2011} allows for realization of a trapping potential of this kind. We set $a_{ho}=\sqrt{\hbar/m\omega}$ as the length unit, where $m$ is the atomic mass. We assume that the trap in the remaining two dimensions is so tight that the dynamics in those directions is frozen. The single-particle Hamiltonian is
\begin{equation}
\frac{H}{\hbar\omega}=-\frac{1}{2}\frac{\partial^2}{\partial x^2}+\frac{1}{2}x^2+\frac{b}{\sqrt{2\pi}d}e^{-x^2/2d^2}.
\label{1ham}
\end{equation}
We diagonalize the Hamiltonian~\eqref{1ham} in the basis of harmonic oscillator states. For high enough barriers, the two lowest lying states become degenerate in energy and separated from the others. One of them corresponds to the even ($\left|\psi_o\right>$) and the other to the odd parity solution ($\left|\psi_e\right>$). Combining them yields two states localized in one of the potential wells: $\left|L\right>=(\left|\psi_e\right>-\left|\psi_o\right>)/\sqrt{2}$ and $\left|R\right>=(\left|\psi_e\right>+\left|\psi_o\right>)/\sqrt{2}$. Reducing to these two modes allows for separation of center-of-mass and relative motion~\cite{Nygaard2008}. Meanwhile, highly excited states of this potential are just harmonic oscillator states. The only important difference between our model potential and a realistic nanoplasmonic trap is the finite depth of the latter one. We thus have to make sure that the states above the realistic trap depth will not be populated during the gate process.

\textit{Feshbach resonance in a double well.}
The crucial component of our proposal is a magnetic Feshbach resonance that couples the pair of atoms with the molecular bound state~\cite{JulienneRMP}. We will describe the resonance using effective two-channel configuration interaction model~\cite{Kokkelmans2002,Julienne2004}. The open channel can be characterized by the hyperfine state of the atomic pair, which we label by $\left|\chi\right>$. The closed molecular channel is denoted as $\left|m\right>$. We neglect the background interaction between the atoms as it gives little contribution close to the resonance. We also treat the weakly bound Feshbach molecule as a pointlike particle of mass $M=2m$. Under these assumptions the Hamiltonian can be written as (see~\cite{Jachymski} for a more detailed derivation)
\begin{equation}
\begin{split}
H=\left|\chi\right>\left<\chi\right|\left(-\frac{1}{2}\frac{\partial^2}{\partial x^2}-\frac{1}{2}\frac{\partial^2}{\partial y^2}+V_{DW}(x)+V_{DW}(y)\right)+ \\ +\left(\left|\chi\right>\left<m\right|+\left|m\right>\left<\chi\right|\right)W(x-y)+\\+\left|m\right>\left<m\right|\left(-\frac{1}{4}\frac{\partial^2}{\partial R^2}+2V_{DW}(R)\right).
\end{split}
\end{equation}
Here $x$ ($y$) is the position of the first (second) particle, $V_{DW}$ is the double well trapping potential and $W$ is the interchannel coupling. For a specific system one should multiply the molecular trapping potential by a factor $\alpha_{mol}/\alpha_{at}$ to account for the change in the polarizability, which determines the trap strength. The general wave function can be written in the form
\begin{equation}
\label{wavef}
\left|\Psi\right>=\left|\chi\right>\sum_{ij}{C_{ij}\psi_i(x)\psi_j(y)}+\left|m\right>\sum_k{A_k\Phi_k(R)},
\end{equation}
where $C_{ij}$ and $A_k$ are the amplitudes and $\psi_i$ are single-particle trap eigenstates with eigenenergies $\epsilon_i$. The molecular wave functions $\Phi_k$ obey the equation
\begin{equation}
\left(-\frac{1}{4}\frac{\partial^2}{\partial R^2}+2V_{DW}(R)\right)\Phi_k(R)=(\mathcal{E}_k +\nu(B))\Phi_k(R),
\end{equation}
where $\nu(B)$ is the energy shift of the state relative to the energy of the open channel. Close to the resonance it may be expanded to first order in magnetic field, giving $\nu=s(B-B_0)$, where $s$ is the difference of magnetic moments between the channel states~\cite{JulienneRMP}.

\begin{figure*}
\centering
\includegraphics[width=0.3\textwidth]{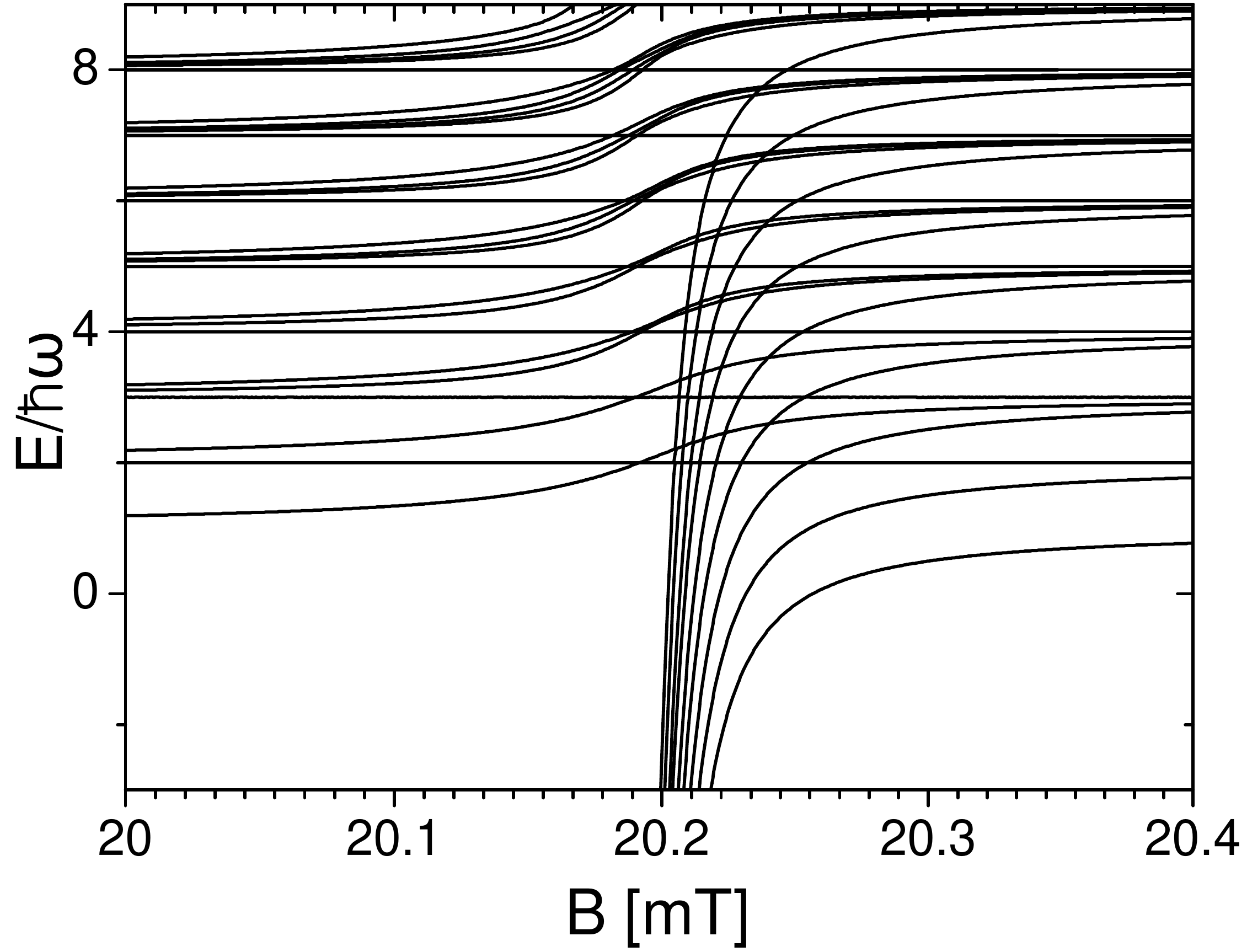}
\includegraphics[width=0.3\textwidth]{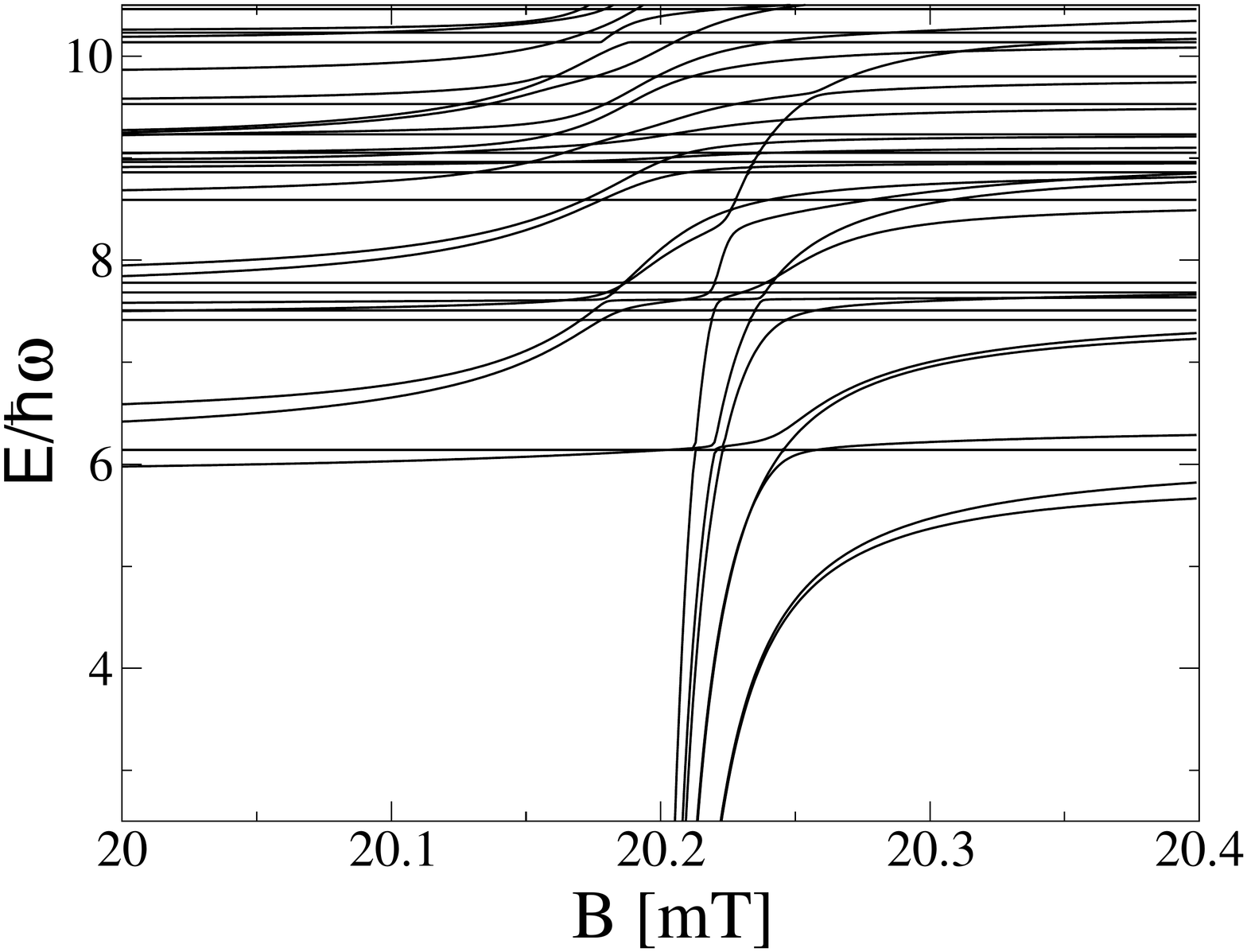}
\includegraphics[width=0.3\textwidth]{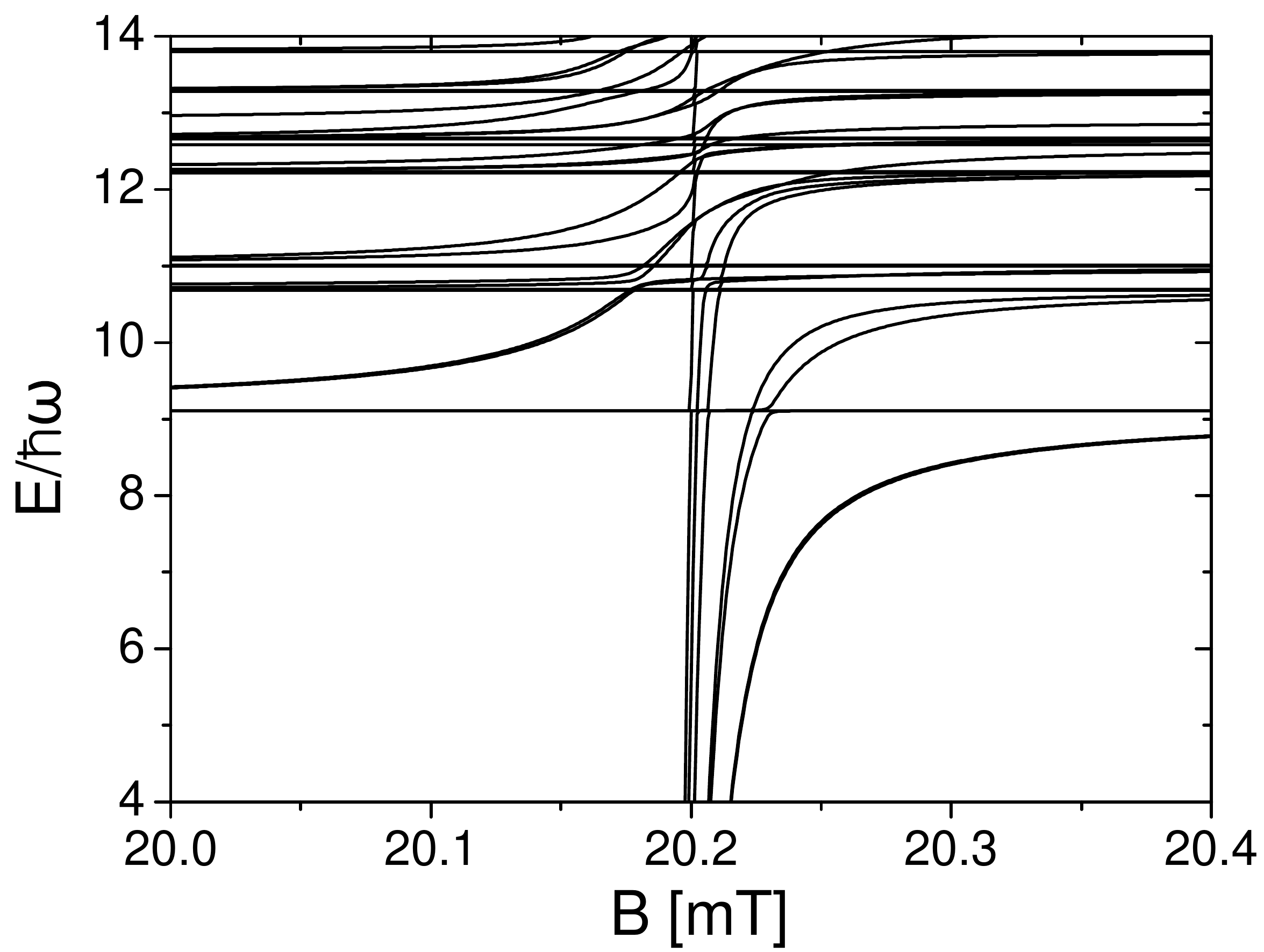}
\caption{\label{fesh}Energy levels for two $^{40}$K atoms in a double well trap in the vicinity of an $s$-wave Feshbach resonance. The harmonic trapping frequency $\omega=10$kHz. The barrier heights are $b=0$ (left), $b=15$ (middle) and $b=100$ (right) and the width $d=1$ in each case.}
\end{figure*}

Substituting~\eqref{wavef} into the Schr\"{o}dinger equation $H\left|\Psi\right>=E\left|\Psi\right>$ leads to a set of equations for the amplitudes
\begin{eqnarray}
(\epsilon_i+\epsilon_j)C_{ij}+\sum_k{V^k _{ij} A_k} = E C_{ij} \\
(\mathcal{E}_k+\nu) A_k+\sum_{ij}{C_{ij}(V^k _{ij})^\star}=E A_k
\end{eqnarray}
where $V^k _{ij}=\alpha\int{dR\,\phi^\star _i(R)\phi^\star _j(R)\Phi_k(R)}$ and $\alpha$ is the coupling constant connected to the resonance parameters~\cite{Jachymski}. By extracting $C_{ij}$ from the first equation, we obtain
\begin{equation}
(E-\mathcal{E}_k-\nu)A_k=\sum_{ijl}{A_l\frac{V^l_{ij}(V^k _{ij})^\star}{E-\epsilon_i-\epsilon_j}}.
\label{finalsum}
\end{equation}
In our treatment the eigenstates of the double well in which we expanded the solution~\eqref{wavef} are superpositions of harmonic oscillator states, so $V^k _{ij}$ can be computed using the matrix elements between them. This can be obtained in terms of hypergeometric functions, and then inserted into Eq.~\eqref{finalsum} to get the energy levels. 

Figure~\ref{fesh} shows the level structure of trapped $^{40}$K atoms near a $202$G $s$-wave Feshbach resonance for three exemplary cases: (i) a~pure harmonic potential, (ii) a~moderate barrier where $b=15$ and $d=1$ and (iii) a higher barrier with $b=100$ with the same $d$. When the barrier is absent, the center of mass and relative motion decouple and the energy spectrum has a simple structure as all the bound states behave in the same way. As the barrier grows, the spectrum becomes more complicated. The two lowest bound levels become close in energy and separate from the rest (see the middle panel). For even higher barriers (right panel), also other bound states form degenerate pairs. In the infinite barrier limit, all the levels are doubly degenerate, which is intuitively clear as the system can be thought of as two completely separated wells. The horizontal lines represent states with odd symmetry, which are not affected by the $s$-wave resonance.

In general, eq.~\eqref{finalsum} may contain divergent terms and requires renormalization~\cite{Jachymski}. However, in one dimension the sums in \eqref{finalsum} converge, in contrast to the three-dimensional case. For the one-dimensional description to be valid, the energy of the particles has to be much smaller than $\hbar \omega_\perp$, where $\omega_\perp$ is the transverse trapping frequency, so that particles occupy the ground state in the transverse direction at each stage of the gate process. Numerical calculations of the gate scheme show that this requires $\omega_\perp \gtrsim 5\omega$. However, even if this is not satisfied, the only complication is that the transversally excited states have to be considered during the evolution. As long as the energy splittings between the different trap states will not constitute the smallest energy scale in the system, this will not affect the speed that the gate can achieve~\cite{Caneva2009}.

\textit{Gate idea.}
To introduce the idea of a quantum gate, it is essential to include the quantum statistics and the spin state of the atomic pair in our considerations. In the case of spin-$1/2$ fermions, the wave function consists of the spatial and spinor part and has to be globally antisymmetric. We identify qubit states with $\left|0\right>=\left|\uparrow\right>$ and $\left|1\right>=\left|\downarrow\right>$. The basis of spin states is $\left|\uparrow\uparrow\right>$, $\left|\downarrow\downarrow\right>$, $(\left|\uparrow\downarrow\right>+\left|\downarrow\uparrow\right>)/\sqrt{2}=\left|\chi_T\right>$ (symmetric) and $(\left|\uparrow\downarrow\right>-\left|\downarrow\uparrow\right>)/\sqrt{2}=\left|\chi_S\right>$ (antisymmetric). Combining the spin states with the trap states in the lowest energy sector~\cite{Hayes2007,Shotter} yields six antisymmetric ground states (in the absence of interactions): $\frac{1}{\sqrt{2}}(\left|LR\right>-\left|RL\right>)\left|\uparrow\uparrow\right>$, $\frac{1}{\sqrt{2}}(\left|LR\right>-\left|RL\right>)\left|\downarrow\downarrow\right>$, $\frac{1}{\sqrt{2}}(\left|LR\right>-\left|RL\right>)\left|\chi_T\right>$, $\frac{1}{\sqrt{2}}(\left|LR\right>+\left|RL\right>)\left|\chi_S\right>$, $\frac{1}{\sqrt{2}}(\left|LL\right>+\left|RR\right>)\left|\chi_S\right>$, $\frac{1}{\sqrt{2}}(\left|LL\right>-\left|RR\right>)\left|\chi_S\right>$.

\begin{figure*}[t]
\centering
\includegraphics[width=0.3\textwidth]{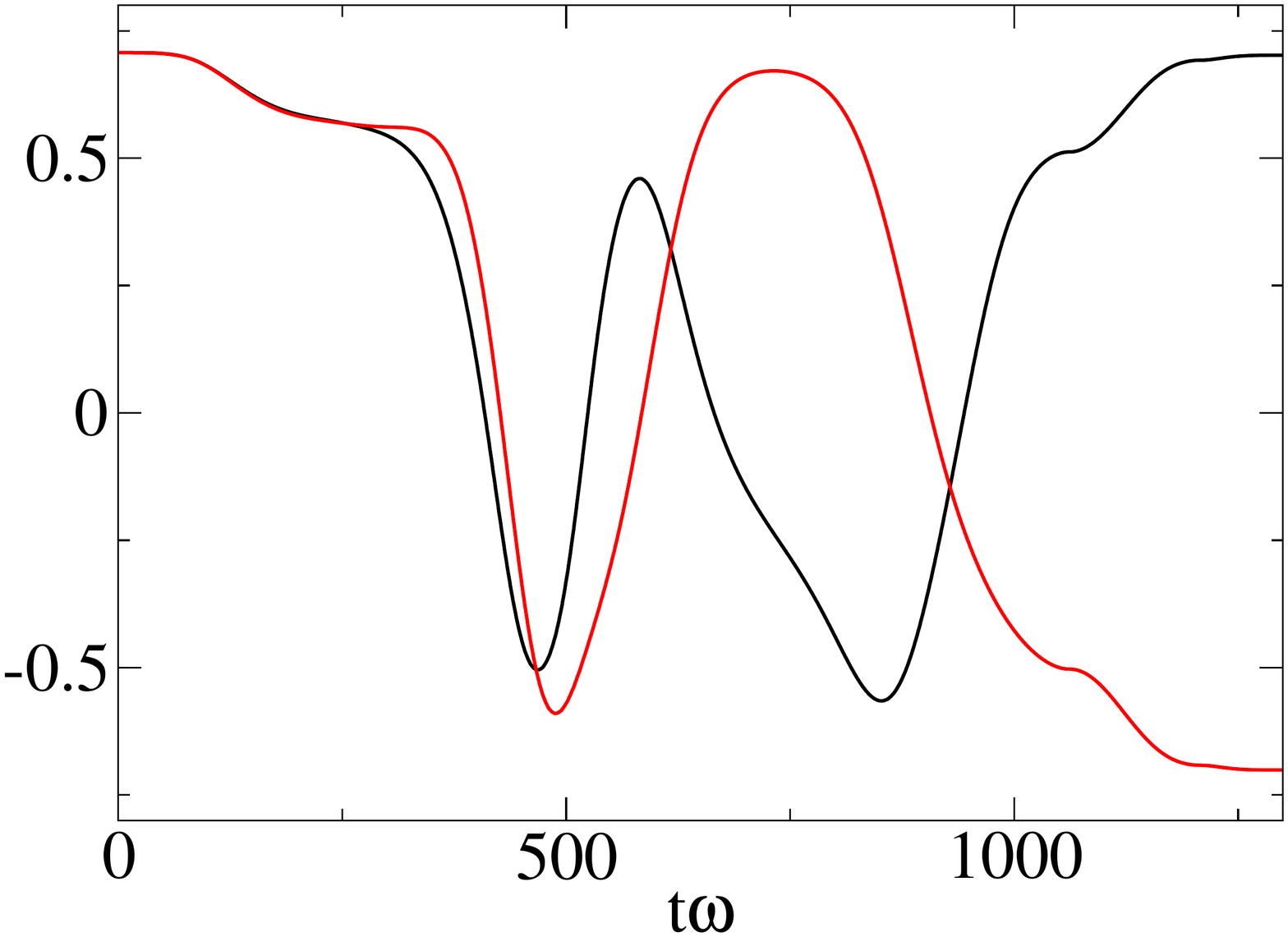}
\includegraphics[width=0.3\textwidth]{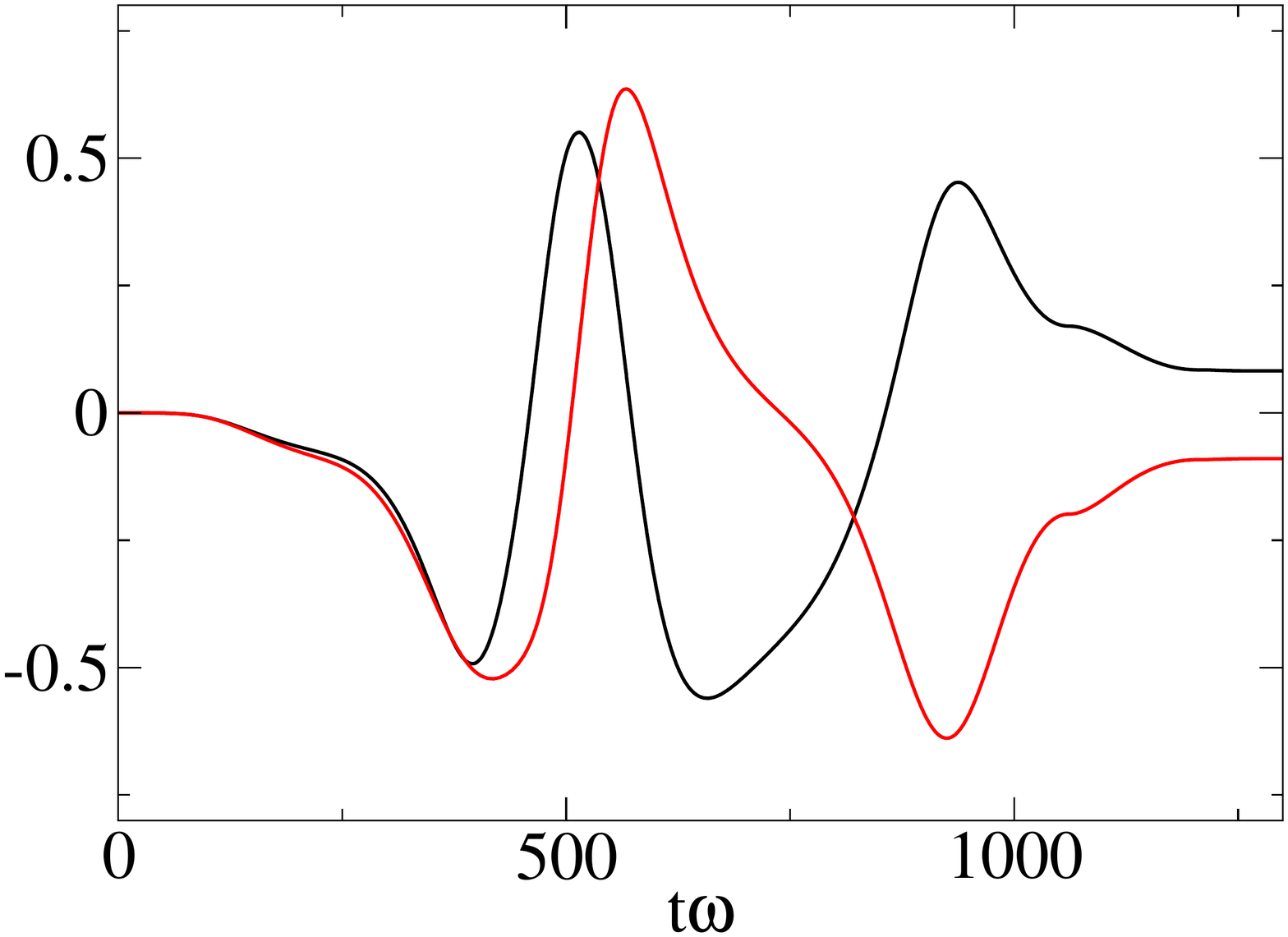}
\includegraphics[width=0.3\textwidth]{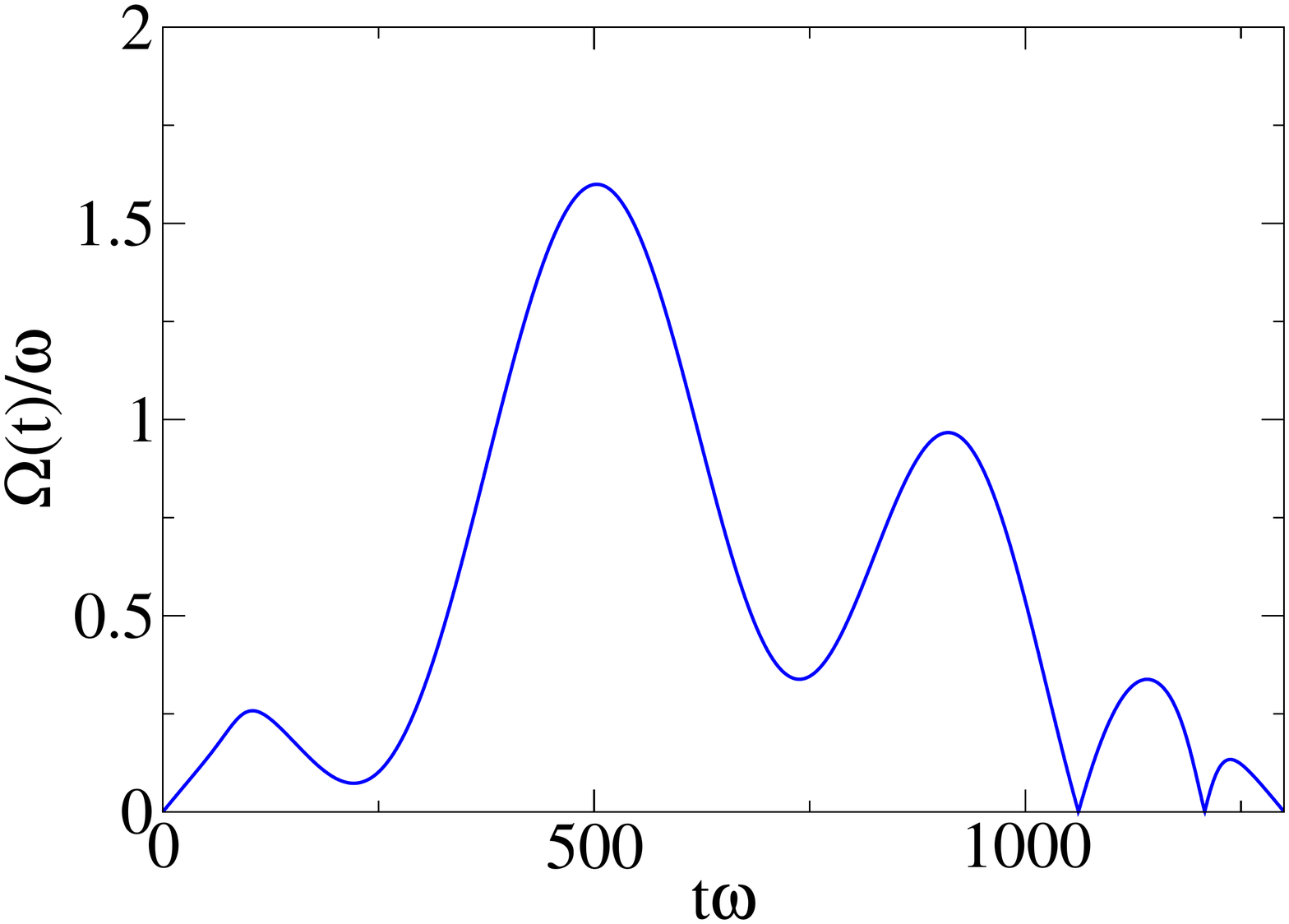}
\caption{\label{fast}(color online) Time evolution of the real (left) and imaginary (middle) parts of the wave function amplitudes of the initial trap states during the optimized gate operation. For the singlet spin state (black lines) the initial spatial part of the wave function is $\frac{1}{\sqrt{2}}(\left|LR\right>+\left|RL\right>)$, while for the triplet state (red lines) we start from $\frac{1}{\sqrt{2}}(\left|LR\right>-\left|RL\right>)$. Right: the optimized pulse shape.}
\end{figure*}

Now let us add the Feshbach resonance to our considerations. We will work with an $s$-wave resonance, where only the singlet spin state is coupled to the molecular channel. The idea of how the gate will work is as follows (see Figure ~\ref{setup}): we start with two fermions in separate wells. An external field is then applied, inducing transition between the ground states and an excited trap state $\left|E\right>$ with  Rabi frequency $\Omega$. Due to the Pauli blocking, which forbids occupation of the same trap state by symmetric spin states, the symmetric and antisymmetric states will evolve differently and accumulate different phases $\phi_s$ and $\phi_a$ during the process. The phase depends only on the state symmetry as the pulse is assumed to couple only to the spatial part of the wave function. A phase gate between states of different symmetry will then be realized when the pair will return to its initial trap state with relative phase $\phi_s-\phi_a=(2n+1)\pi$, $n\in\mathbb{N}$. In the computational basis this results in a SWAP gate. An entangling gate is realized if we choose the phase to be equal to $(n+1/2)\pi$. The truth table for this operation is $\left|00\right>\rightarrow\left|00\right>$, $\left|11\right>\rightarrow\left|11\right>$, $\left|10\right>\rightarrow \frac{1}{\sqrt{2}}(e^{i\pi/4}\left|10\right>+e^{-i\pi/4}\left|01\right>)$, $\left|01\right>\rightarrow \frac{1}{\sqrt{2}}(e^{i\pi/4}\left|01\right>+e^{-i\pi/4}\left|10\right>)$. This gate belongs to the $\sqrt{SWAP}$ universality class and allows for universal quantum computation~\cite{Whaley,Muller}. The state readout can be achieved using subwavelength imaging techniques~\cite{Gorshkov2008}. One potential source of errors is that for the singlet spin state the particles can end up in the same potential well. This problem is avoided by the resonance, which shifts the energy of such state and thus decouples it from the transitions. By analyzing Figure~\ref{fesh} one can conclude that the optimal choice is to work at fields slightly lower than the position of the resonance, where the energy shift of the bound levels is large. In this case particles in the same well form a far detuned bound state.

Coupling the trap states can be achieved using Raman transitions. It could be desirable to couple only to a single target state $\left|E\right>$, chosen such that it is energetically separated from other trap states and the Franck-Condon factors $\left<E\right|e^{ikx}\left|R\right>$, $\left<E\right|e^{ikx}\left|L\right>$ (denoted by $\eta_{RE}$ and $\eta_{LE}$) are large. However, other trap states will also unavoidably get populated during the gate process, unless the ratio $\Omega/\omega \ll 1$ (equivalent to the Lamb-Dicke regime~\cite{Leibfried2003}). Keeping $\Omega$ low will result in long operation times, so one can expect that a shaped pulse will be needed to operate at high Rabi frequencies while maintaining high fidelity.

\textit{Implementation and optimization of the gate.}
We will now consider more specifically a gate implementation with $^{40}$K atoms. We will work at barrier height $b=36$ and width $w=1.5$ in oscillator units. This choice of trap parameters gives two almost degenerate lowest states, while the first excited state is separated from the next one. In the high barrier regime the Franck-Condon factors are equal for the states localized in left and right potential wells. When trying to find optimal and realistic experimental parameters which will give shortest operation times, one finds two tradeoffs. Firstly, higher Rabi frequencies lead to faster dynamics but also introduce losses via leakage to highly excited states. Secondly, large trapping frequencies reduce the characteristic timescales, but at the same time lower the Franck-Condon factors as the trap becomes smaller and the laser wavelength cannot. Thus the time needed for the operation does not scale linearly with the trapping frequency. For the Raman transition the achievable wavenumbers are of the order of $0.03$nm$^{-1}$. Using UV transitions would allow to improve it by around $50\%$, but we will assume optical transitions which are far more convenient experimentally. 

The target state after the operation is to have particles again in the lowest trap states ($\left|LR\right>\pm\left|RL\right>$), but with relative phase $\pi$ between the symmetric and antisymmetric spin states. The fidelity of the gate can be defined as $f=\left|\left<\psi_{out}\right|\left.\psi_{target}\right>\right|^2$. We choose the initial pulse to have the form $\Omega(t)=\Omega_0 t/\tau(1-t/\tau)$, where $\tau$ is the operation time. We set the trapping frequency $\omega=2\pi\times 5$ MHz for which the well minima are $\sim 15$nm apart and $\tau=1300/\omega\approx 42$~$\mu$s. To optimize the pulse and achieve high fidelity we applied the CRAB optimization method~\cite{CRAB}. This algorithm starts with an initial pulse and seeks an optimal correction written in terms of truncated Fourier series with randomized frequencies. After optimization the fidelity of the solution reached over $99.9\%$. The gate operation corresponding to it is depicted on Figure~\ref{fast}. The Rabi frequency of the pulse $\Omega$ does not exceed $1.8\,\omega\approx 2\pi\times 9$ MHz, which is a reasonable value for $^{40}$K~\cite{Ospelkaus2008}. The obtained gate time is limited by the value of $\Omega$ and does not reach the quantum speed limit~\cite{Caneva2009}. However, using stronger pulses makes the leakage effects stronger and would require adding more driving fields.  

\textit{Conclusions.}
We proposed a new scheme for realizing a quantum gate with ultracold atoms which does not need access to individual particles. We discussed the implementation of our method in a tight nanoplasmonic trap for realistic experimental conditions and showed that it is possible to obtain a high fidelity gate with operation times considerably shorter than in previous proposals. Our scheme does not crucially depend on the details of the trapping potential or internal structure of the atoms. Quantitative corrections to energy levels originating from trapping and interaction potential details would be important in practical implementation, but can be taken into account or tackled by means of closed-loop optimal control~\cite{Rosi2013}. While solid state gates operate on sub-$\mu$s timescales~\cite{Sank2014}, predicted operation to decoherence time ratio in such systems is of the order of $10^{-3}$ -~to be compared against roughly $10^{-4}$ here. This makes our proposal a competitive candidate for the realization of quantum processors.

We would like to thank Peter Zoller and Darrick Chang for stimulating discussions. The research leading to these results has received funding from the European Union's Seventh Framework Programme (FP7/2007-2013) under  grant agreement No. 600645 - IP SIQS. This work was supported by the Foundation for Polish Science International PhD Projects Programme co-financed by the EU European Regional Development Fund, SFB/TRR21 and National Center for Science Grant No. DEC-2011/01/B/ST2/02030 and DEC-2013/09/N/ST2/02188.

\bibliography{allrefs}
\end{document}